%% file: main.tex
\documentclass[conference]{IEEEtran}

\usepackage{cite}
\usepackage{amsmath,amssymb,amsfonts}
\usepackage{algorithmic}
\usepackage{graphicx}
\usepackage{textcomp}
\usepackage{xcolor}
\usepackage{tabularx}
\def\BibTeX{{\rm B\kern-.05em{\sc i\kern-.025em b}\kern-.08em
    T\kern-.1667em\lower.7ex\hbox{E}\kern-.125emX}}

\begin{document}

\title{Handling Communication via APIs for Microservices}

 \author{\IEEEauthorblockN{Vini Kanvar}
 \IEEEauthorblockA{ \textit{IBM Research},
 India\\
 vkanv031@in.ibm.com}
 \and
 \IEEEauthorblockN{Ridhi Jain}
 \IEEEauthorblockA{\textit{IIIT Delhi},
 India \\
 ridhij@iiitd.ac.in}
 \and
 \IEEEauthorblockN{Srikanth Tamilselvam}
 \IEEEauthorblockA{\textit{IBM Research},
 India\\
 srkanth.tamilselvam@in.ibm.com}
 }

\maketitle

\input{Sections/abstract}

\input{Sections/introduction}
\input{Sections/challenges}

\input{Sections/json}
\input{Sections/communication}

\input{Sections/methodology}
\input{Sections/exposingApis}

\input{Sections/discussion}
\input{Sections/experiments}
\input{Sections/study}
\input{Sections/related}
\input{Sections/conclusion}

\bibliographystyle{plain}
\bibliography{main}

\end{document}

%% file: Sections/abstract.tex
\begin{abstract}
\label{abstract}
Enterprises in their journey to the cloud, want to decompose their monolith applications into microservices to maximize cloud benefits. Current research focuses a lot on how to partition the monolith into smaller clusters that perform well across standard metrics like coupling, cohesion, etc. 
However, there is little research done on taking the partitions, identifying their dependencies between the microservices, exploring ways to further reduce the dependencies, and making appropriate code changes to enable robust communication without modifying the application behaviour. 

In this work, we discuss the challenges with the conventional techniques of communication using JSON and propose an alternative way of ID-passing via APIs. We also devise an algorithm to reduce the number of APIs. For this, we construct subgraphs of methods and their associated variables in each class and relocate them to their more functionally aligned microservices. Our quantitative and qualitative studies on five public Java applications clearly demonstrate that our refactored microservices using ID have decidedly better time and memory complexities than JSON. Our automation reduces 40-60\% of the manual refactoring efforts.
\footnote{This paper is an extended version of our paper~\cite{icse23} published in ICSE'23 NIER https://ieeexplore.ieee.org/document/10173907.}

\end{abstract}

%% file: Sections/introduction.tex
\section{Introduction}
\label{introduction}


Monolith architecture encapsulates all capabilities into a single deployable unit. It offers benefits like onboarding developers quickly and making deployment easier~\cite{taibi2017microservices}. However, maintaining monolithic applications gets difficult as they age, as developers find it difficult to predict or isolate the change impact. Therefore, microservices architecture is seen as an alternative. Microservices advocate building an application as a connection of simple services, each performing a single business function. Microservices offer many benefits such as improved fault tolerance, flexibility to choose different technology, the ability to organize teams as per functionality, etc. They are also best suited for cloud deployment due to the utilization and cost benefits associated with deploying and scaling services individually \cite{learn_micro}. Therefore, enterprises are increasingly refactoring their legacy monolith application into microservices as part of their journey to cloud~\cite{dasgupta2021ai, kalia2020mono2micro}.

Although there are several works on creating microservices via APIs, there is little work on how data should be communicated via the APIs. 
The communication via APIs is non-trivial given that the memory address space and resources are not shared between microservices. Existing literature proposes the use of JSON format for data communication. Data communication in JSON format includes the transfer of JSON data from client to server~\footnote{Two microservices act as a client and a server if one calls the APIs of the other~\cite{icws22}. A microservice can act as both a client and a server with respect to different APIs.}, reconstruction of objects at the server, returning to the client any updates made in the API by the server, and reconstruction at the client.

When microservices are written from scratch, APIs are usually made stateless~\cite{icws22}. In other words, objects do not contain complex constructs like pointers, static fields, or external resources, and their object constructors are simplistic. However, objects include these constructs when we refactor a monolith to microservices. Serialization to JSON format loses information in the presence of these constructs. Existing literature~\cite{icws22} demonstrates how transfer of objects that contain pointers may lead to information loss. Nonetheless, it neither highlights nor solves the plethora of other issues in using JSON format for communication. 

\begin{figure*}
\centering
\includegraphics[width=180mm]{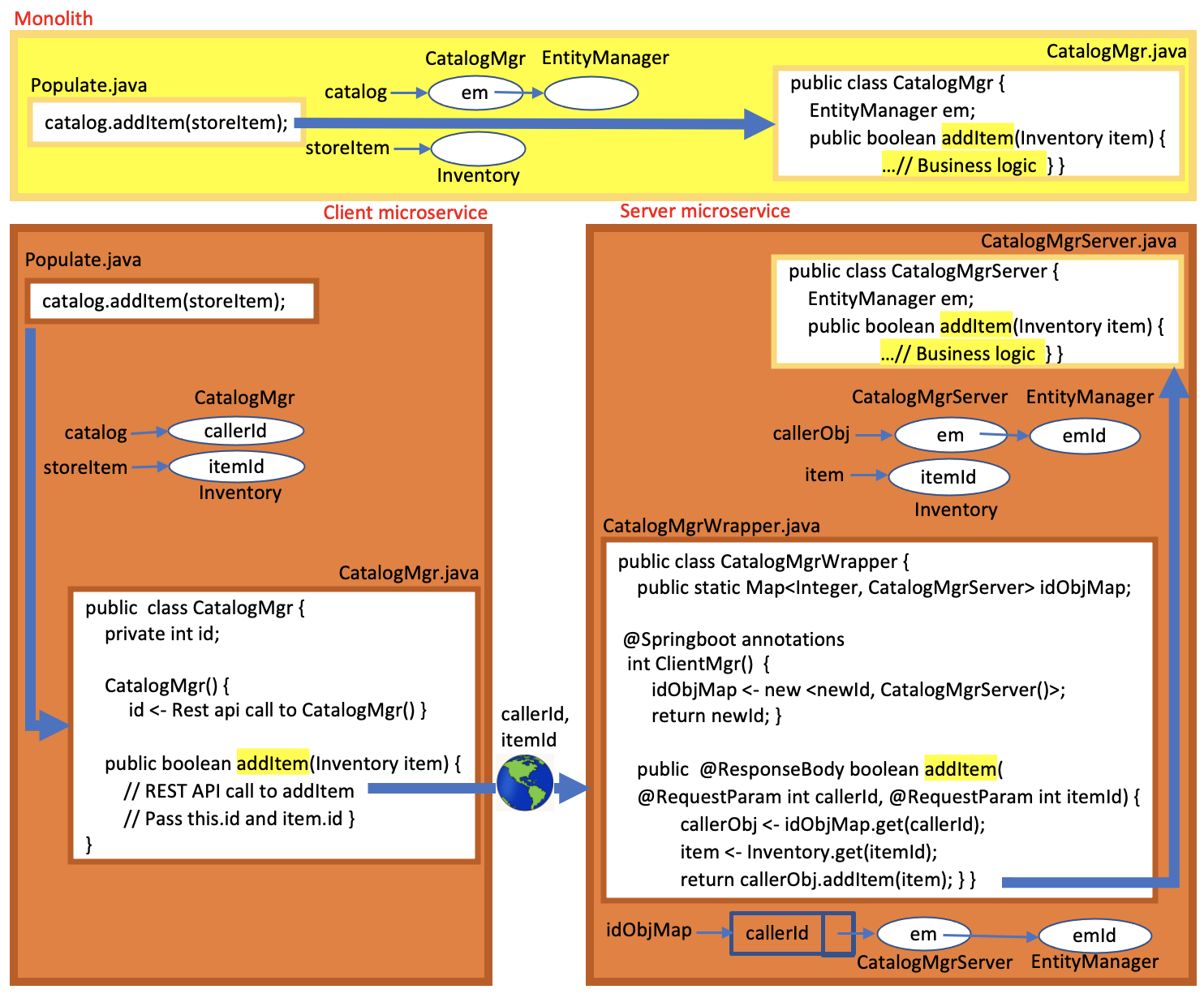} 
\caption{Example to motivate the need of using ID-passing approach in communication between microservices. Original Plantsbywebsphere monolith communicates complex non-primitive objects. Its refactored microservices (output from our tool) need to communicate the objects using their IDs. Detailed code is in Figure~\ref{fig:refactored}.}
\label{fig:id}
\end{figure*}

\begin{figure*}
\centering
\includegraphics[width=180mm]{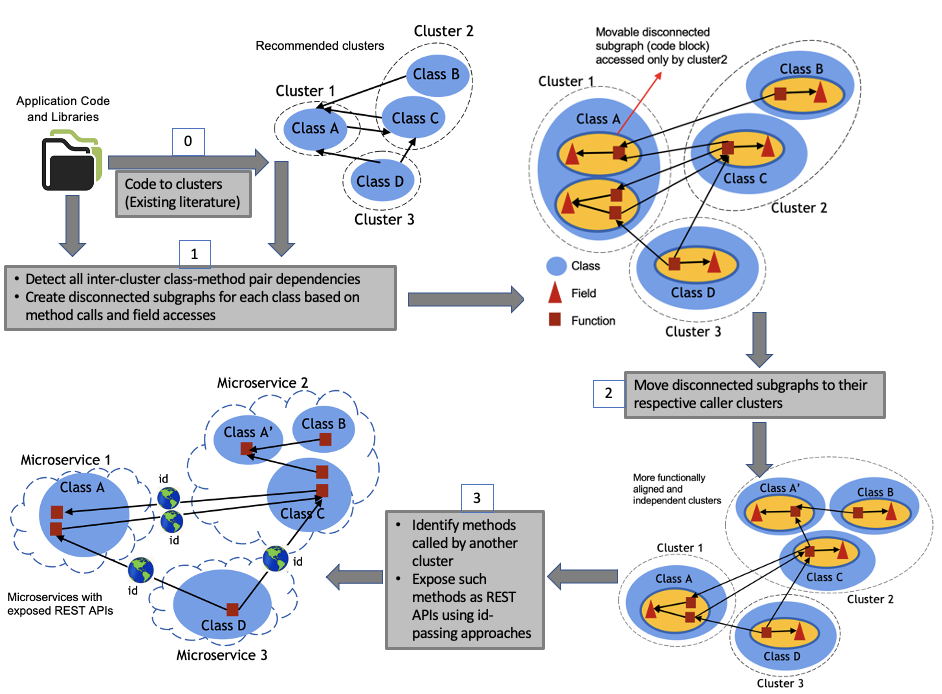}
\caption{System Diagram: Clusters to microservices (steps 1, 2, 3). Code to clusters is done by existing literature (step 0).}
\label{system}
\end{figure*}


Section~\ref{json} demonstrates and categorizes issues in JSON transfer in detail. Some issues are the following: If an object containing static fields is converted to JSON format, the values of the static fields are lost. Further, the JSON format does not capture aliases in objects. Also, deserialization from JSON to an object requires calling the object's constructor. However, constructors are called blindly by JSON libraries~\cite{springframework}; for instance, they violate any singleton property imposed by the programmer and violate any private access modifier.

We propose the use of IDs for data communication. Here we do not transfer the object between client and server. Instead, we transfer a globally unique ID of the object assigned to it by the microservice that created it. The map of the ID to the object is maintained by the microservice that created it. The ID is transferred in a wrapper class of the same name to minimize the code refactoring from monolith to microservices. The object is fetched from the map whenever its values need to be read or written.

Figure~\ref{fig:id} is an example to motivate the need to pass ID instead of JSON. It is a snapshot from an open-source application, namely, PlantsByWebsphere (PBW)~\cite{pbw}. PBW is a monolith application, i.e., a single deployable unit, for purchasing various plants. We show how to refactor it to microservices that communicate via APIs and transfer data using IDs. In the figure, arrows between files represent function calls. Graphs shown beside these arrows represent memory/objects being transferred via the functions, oval nodes represent memory locations, and edges between oval nodes represent memory links or pointers to the target location. Identifiers like CatalogMgr and Inventory, shown above or below the oval nodes, denote the class type of the memory location. Identifiers like em, shown inside the oval nodes, represent member fields of the class. Identifiers like catalog and storeItem, pointing to the oval nodes, denote variables. In the monolith application (shown on the top of the figure, in yellow), class Populate calls method addItem(.) of class CatalogMgr. The call is made using object catalog and parameter object storeItem of class Inventory. 

Using existing clustering techniques~\cite{desai2021graph}, class Populate goes into one microservice, and class CatalogMgr goes into another due to their functional properties. Therefore, method addItem(.) needs to be exposed as an API, and objects catalog and storeItem need to be communicated via the API. 
JSON format is often used for the communication of data. However, class Inventory contains several private and static member fields (not shown in the figure). Both classes, Inventory and CatalogMgr, contain pointers. Therefore, serialization of these objects to JSON loses values of static fields, exposes values of private fields, and loses alias information.

Figure~\ref{fig:id} shows our refactoring of the monolith into microservices in orange. Here we propose to communicate data using their IDs. Therefore, we assign IDs to objects catalog and storeItem at the time of their creation. We pass their IDs, namely, callerId and itemId, shown between the microservices in the figure. In order to minimize the refactoring, we do not modify the original classes Populate and CatalogMgr. Instead, we add a class CatalogMgr to the client microservice. It acts as a wrapper class and contains the unique ID, and makes the REST API call to addItem(.). In the call to addItem(.), ID is communicated from Populate class to the wrapper class CatalogMgr in the client microservice using an object of this wrapper class. We add a class CatalogMgrWrapper to the server microservice. It acts as a wrapper class, contains the id-to-object map, and exposes addItem(.) as a REST API. Whenever the object members need to be read or written, the object is fetched from the map using its ID. In the server microservice, since the caller object is required to call its member function addItem(.), the function of CatalogMgrWrapper fetches the object callerObj from the map using callerId. It then makes a call to the original addItem(.), which contains the original business logic. 





In this work, we also propose a technique to reduce communication between microservices. 
To reduce communication, we need to reduce the number of API calls between the microservices. For this, we propose to relocate methods between microservices. We call this technique as {\em function isolation}, which uses disconnected subgraphs of methods and fields.

The rest of the paper is organized as follows. We discuss our system diagram in Section~\ref{sec:introsys}. Section~\ref{challenges} presents our key contributions and challenges in code refactoring. We describe the issues in using JSON for data transfer via APIs in Section~\ref{json}. In section~\ref{communication}, we propose an ID-passing approach between microservices to retain the behaviour of the monolith. Section~\ref{methodology}, we propose the technique to reduce communication between microservices. Generation of REST API calls using Springboot framework are explained in Section~\ref{exposingapis}. Coding design considerations for microservices are highlighted in Section~\ref{discuss}. Experimental setup and results on five applications are discussed in detail in Section~\ref{experiments}. Implementation and results of the experiments for the five applications are available on Github~\cite{res}. Section~\ref{experiments} presents experiments and discusses our qualitative study with developers to know the usefulness of our output. Section~\ref{related} presents the related work and Section~\ref{conclusion} concludes the paper with future work.

\section{System Diagram}
\label{sec:introsys}

Figure~\ref{system} shows our system diagram, which starts by converting a monolith into clusters of functionally aligned classes. This is shown as step 0 and is a well-studied work in literature~\cite{desai2021graph, mitchell2006automatic, jin2019service, kalia2020mono2micro, mathai2021monolith}. The clustering is performed with an optimization objective to produce high cohesion within the cluster and low coupling across clusters.
The clusters are represented as graphs with classes as nodes and relationships like the method calls as edges. In order to deploy these clusters, the inter-cluster dependencies (edges) need to be resolved, i.e., the methods need to be exposed as APIs, method calls need to be converted to REST API calls, and data needs to be communicated via IDs (as proposed in the introduction section).
We propose to perform the following steps:
\begin{itemize}
    \item Function isolation: Relocate methods to their more functionally dependent clusters wherever applicable to reduce the inter-cluster dependencies, thereby reducing data communication between the microservices. (Steps 1 and 2 of Figure~\ref{system}).
    \item Transfer data using IDs between microservices via APIs (Step 3 of Figure~\ref{system}). 
\end{itemize}

In step 0 of Figure~\ref{system}, we use existing work to group classes in the application as functional clusters. In step 1, we create disconnected subgraphs of methods and fields. In step 2, disconnected subgraphs that are accessed by only one other cluster are relocated to the calling cluster to reduce the overall cross-cluster dependencies, thus, reducing data communication between microservices. For example, the top subgraph of class {\em A} is relocated as class {\em A'} to cluster 2 because it is disconnected and called only by cluster 2. In step 3, methods that cannot be relocated and are accessed from different clusters are exposed as APIs. These APIs enable communication between the clusters deployed in different cloud containers/virtual machines. Unlike monolith, address space and resources are not shared between microservices. Thus, we propose to pass objects between microservices using their unique IDs to preserve the behaviour of the monolith. 

%% file: Sections/challenges.tex
\section{Key Contributions and Challenges}
\label{challenges}
The key contributions of our paper are :
\begin{itemize}
\item Devising a technique to share objects using IDs between microservices. Note that memory address space and resources are not shared, and JSON loses information.
\item Refactoring in the least intrusive way, i.e., new code in the microservices does not mix with the original business logic.
\item Reducing data communication between microservices.
\item Creating more functionally aligned microservices.
\end{itemize}



Several tools recommend which classes should be part of which candidate microservice~\cite{kalia2020mono2micro, mitchell2006automatic, jin2019service, DBLP:conf/aaai/DesaiBT21}. However, refactoring of the code is still required to enable communication between microservices. We address the following key challenges.

\begin{itemize}
\item {\em Exposing REST APIs}. API rewriting can be a labor-intensive task. Automation should cover (i) converting method calls to REST API calls and (ii) changing parameters and return types to a format i.e. IDs that can be exchanged between microservices. Exchanging non-primitives via APIs should preserve properties like complex data structures, including recursive references, static member fields, singleton property, state of parameters and caller object in the API, static resolution of dynamic API dispatch, and others. These are lost in serialization to JSON.

\item {\em Clean Refactoring}. Modifications in the monolith code should be made such that it is not all mixed up with the monolithic code, and developer can perform dedicated development for exposing REST APIs.

\item {\em Function isolation may introduce new errors}. Relocating methods and their fields may cause errors like missing imports and missing fields. We refactor considering these errors.


\item {\em Optimal solution may take several iterations}. Every pass of code relocation may uncover opportunities to further eliminate newer dependencies that were not possible earlier. An upper bound on the number of iterations is non-deterministic. Our empirical measurements are for a single iteration, but our approach can be applied over multiple iterations.
\end{itemize}


In the following section, we identify and categorize the issues in using JSON format for data communication via APIs between microservices.

%% file: Sections/json.tex

\begin{figure*}
  \centering
  \begin{tabularx}{\textwidth}{X}
    \includegraphics[width=178mm]{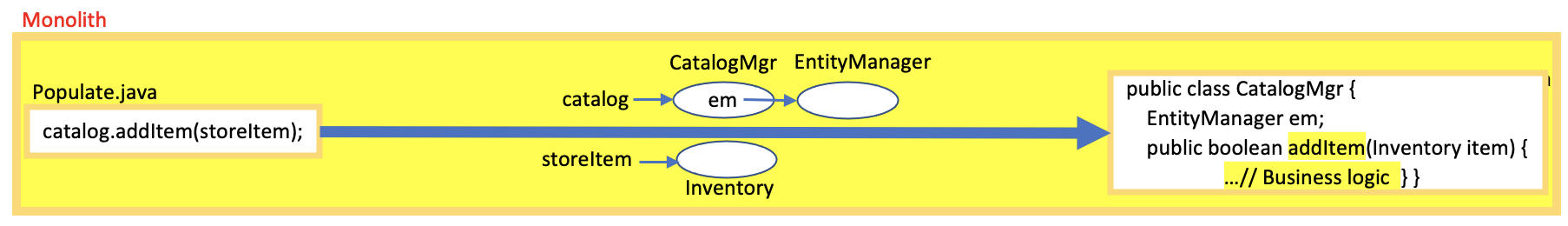} \\
    {\bf \small (a) Monolith Design containing classes Populate and CatalogMgr. Method addItem(.) is defined in class CatalogMgr and called by class Populate.}
    \\
    \includegraphics[width=178mm]{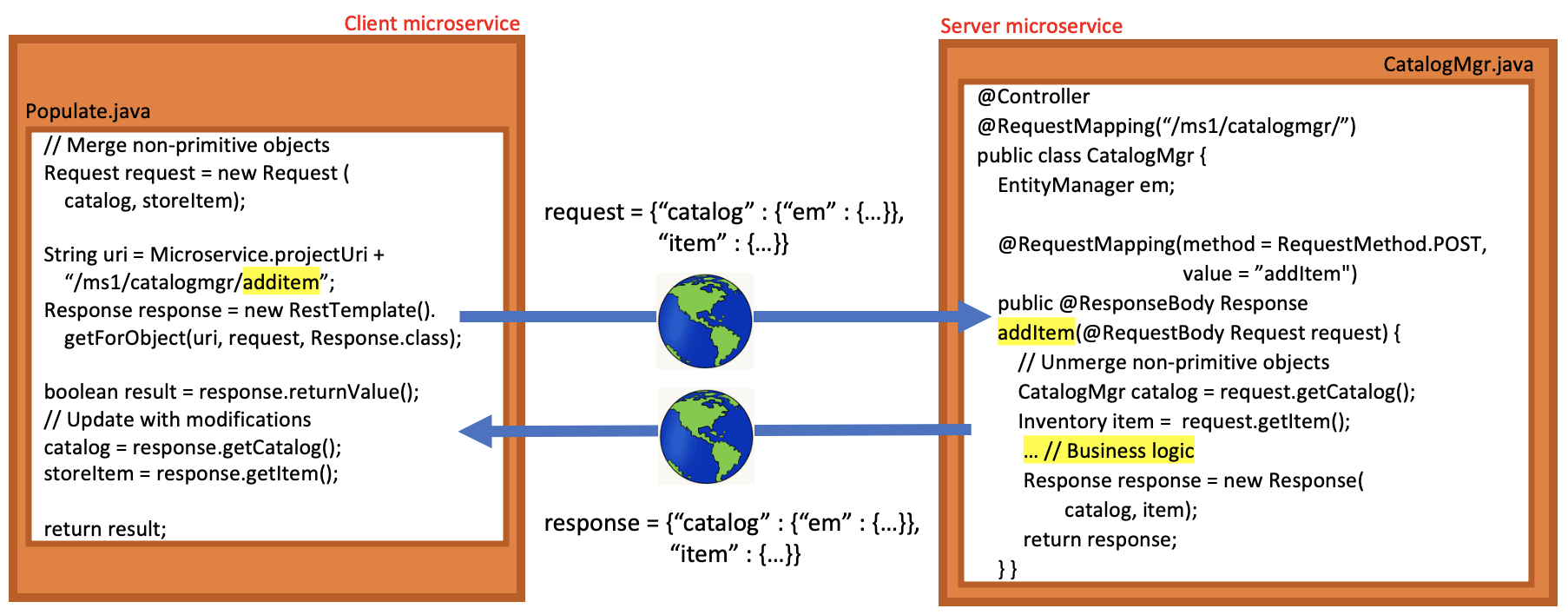} \\
    {\bf \small (b) Microservices with class Populate and class CatalogMgr, respectively. Method addItem(.) exposed as API. Call to addItem(.) refactored to API call. The client and server communicate in JSON format using request and response.}
  \end{tabularx}
  \caption{Monolith application refactored to microservices design showing the code changes without using the ID-approach.}
  \label{wrapper}
\end{figure*}

\section{JSON for Data Transfer}
\label{json}

Monolith assumes a shared memory address space and resources between methods. However, the memory address spaces and resources are separated when converting the monolith to a microservices-based architecture. As a result, Java object references cannot be shared; and thus, an effective technique to share non-primitive objects is required. A client needs to pass the caller object and arguments. A server must return objects and reflect any state changes made in the arguments to the client.

Conventionally, objects are shared via APIs in JSON format.  Section~\ref{sec:refactor} explains how to refactor the monolith code to communicate data in JSON format. Section~\ref{existing} discusses the different issues that arise in using JSON format for communication.

\subsection{Refactoring using JSON for Data Transfer}
\label{sec:refactor}
Figure~\ref{wrapper} shows how to refactor the monolith code into microservices where data is communicated via APIs in JSON format. We continue with our motivating example of Section~\ref{introduction}. Figure~\ref{wrapper}(a) shows the monolith code where addItem(.) is called using object catalog and argument storeItem. The method addItem(.) is defined in class CatalogMgr. Figure~\ref{wrapper}(b) shows the refactoring into microservices. Existing clustering techniques~\cite{desai2021graph} create a microservice containing class Populate and another microservice containing class CatalogMgr. Method addItem(.) needs to be exposed as an API since it is called outside the microservice. The microservice that exposes the API is called the server microservice, and the one that calls the API is called the client microservice. 

In the client microservice, non-primitive objects catalog and storeItem are stored in the variable \textit{request}. A REST API call is made, and \textit{request} is passed in JSON format, as shown in the figure. The server microservice deserializes the received \textit{request} into objects catalog and item. After executing the business logic of addItem(.), the updated catalog and item are stored in variable response. The response is returned in JSON format, as shown in the figure. The client microservice deserializes the received response into objects catalog and storeItem so that any state update done by the server is reflected in the client.

\begin{figure*}
    \centering \small
\begin{tabular}{@{}ll@{}}
\begin{tabular}{@{}c@{}}
\includegraphics[width=95mm]{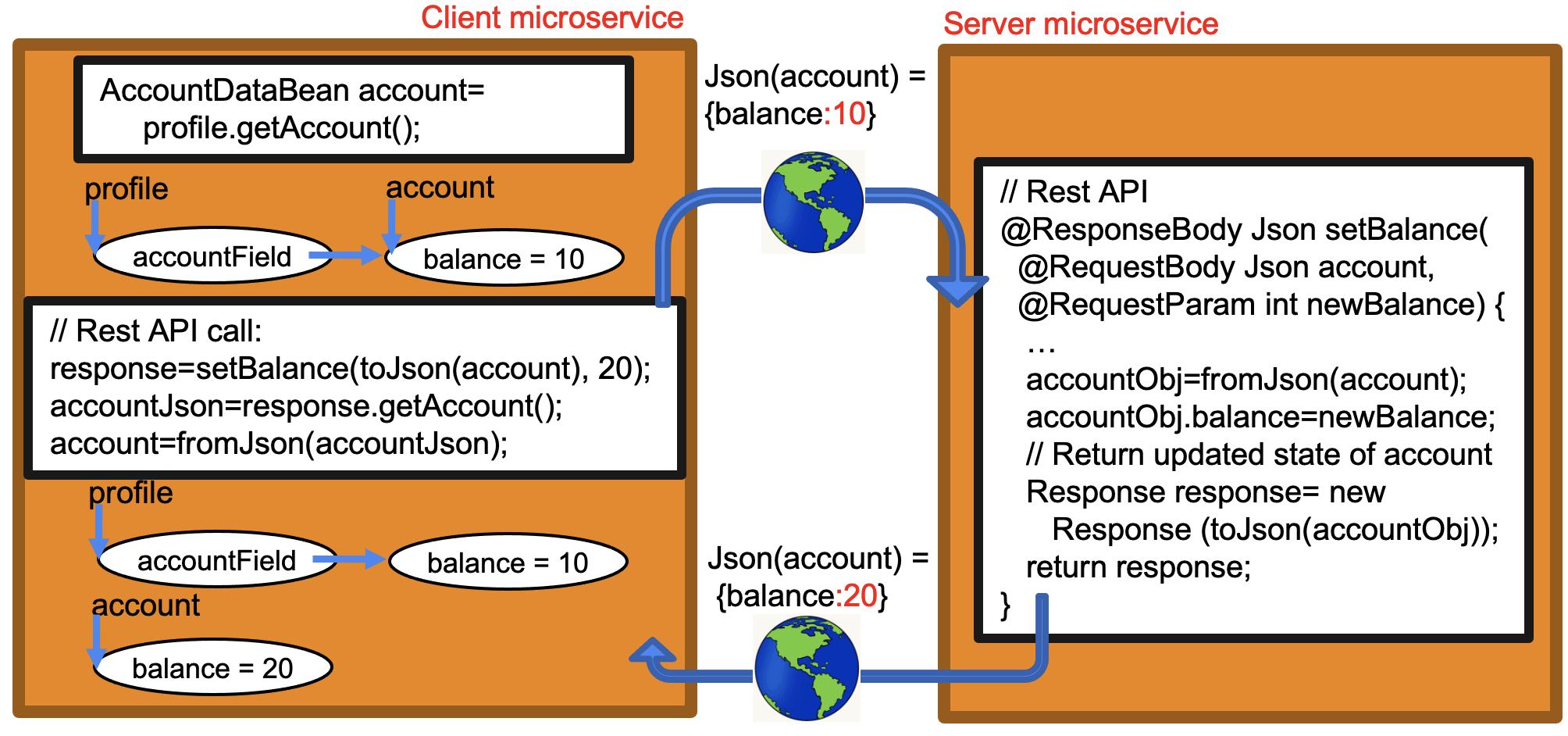} \\
\footnotesize (a) References get lost after deserialization \\ \hline
\includegraphics[width=95mm]{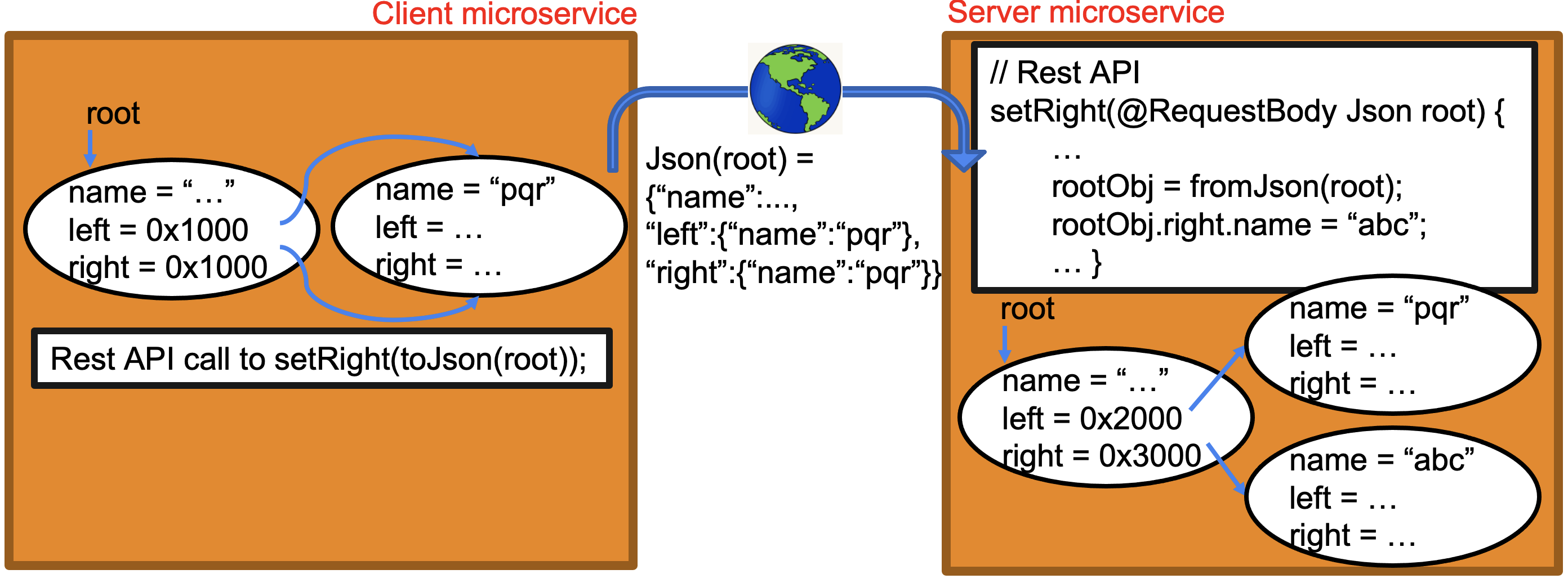} \\
\footnotesize (b) Aliases are lost in serialization to JSON\\ \hline 
\end{tabular}
&
\begin{tabular}{@{}c@{}}
\includegraphics[width=70mm]{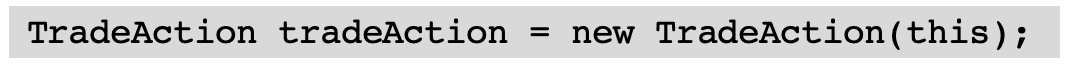} \\
\footnotesize (c) 'this' cannot be updated  \\  \hline
\includegraphics[width=75mm]{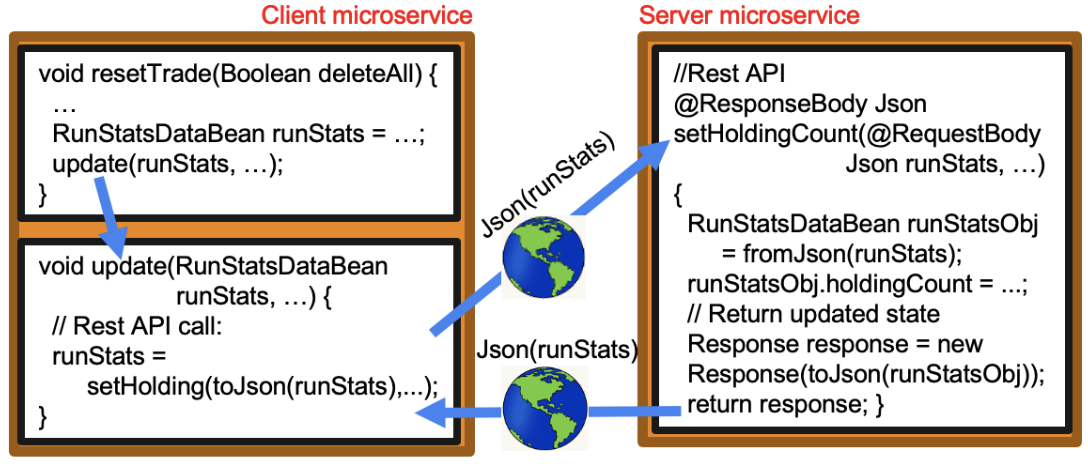} \\
\footnotesize (d) State of a modified object may not be reflected to the client \\ \hline 
\includegraphics[width=75mm]{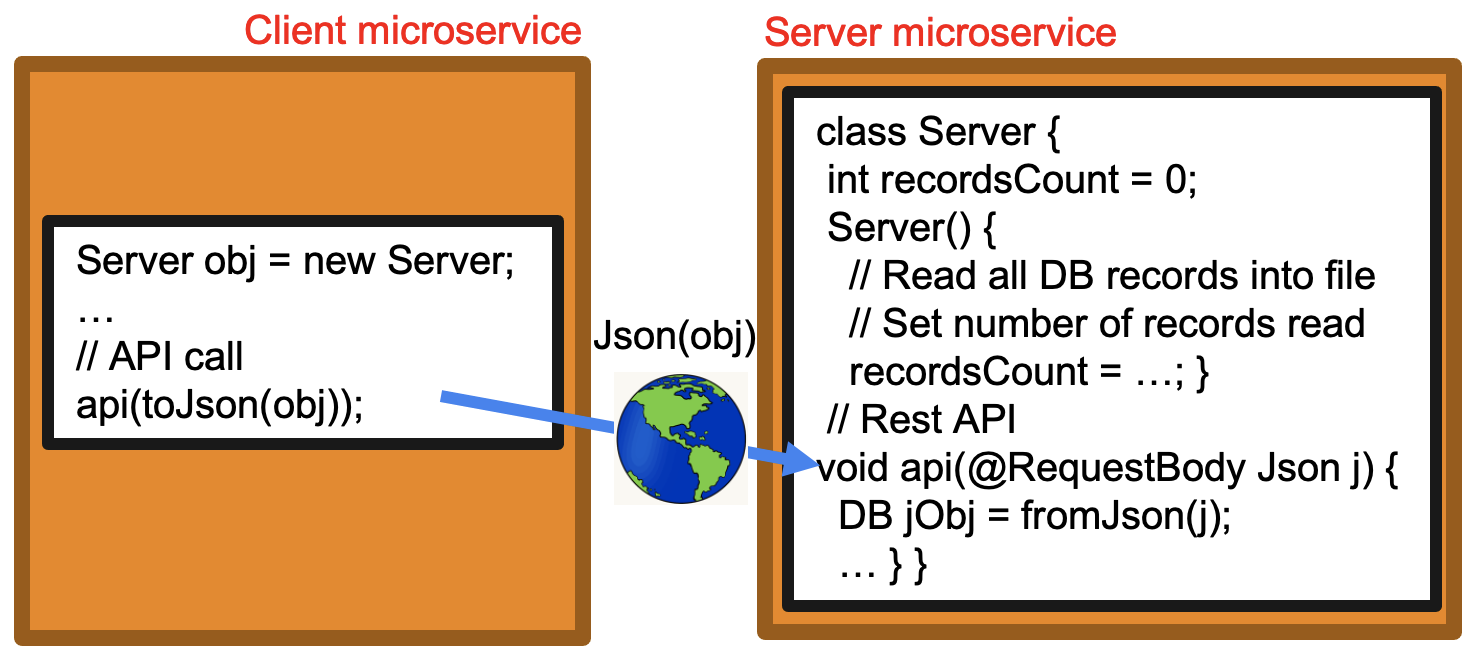} \\ 
\footnotesize (e) Constructor is called assuming non-default field values \\ \hline
\end{tabular}
\end{tabular}
\setlength{\abovecaptionskip}{0pt}
\caption{Problems in passing objects between microservices in JSON format.}
\label{jsonproblems}
\end{figure*}

\subsection{Issues with JSON for Data Transfer}
\label{existing}
Usually, in REST services, data is exchanged in JSON format. The client serializes data into JSON, passes it, and the server then deserializes the received JSON object. But in the context of an automated monolith to microservices migration, simply converting a Java object into JSON object can lead to erroneous behavior of the application. Below we list categories of cases where passing objects in JSON format via APIs leads to loss of semantics, resulting in erroneous application behavior.


JSON {\em  issues related to access modifiers and data types:}
\begin{itemize}
\item Static field values are lost. For instance, in PBW, class CatalogMgr has 
several static member fields. JSON of an object does not save the value of static member fields since they are not saved in the object but in the global address space, i.e., class definition. Therefore, whenever an object of class CatalogMgr 
is passed via argument or returned between microservices, its static field values are lost.
\item File handler and socket IDs cannot be passed between microservices. If java.io.File is a member field, JSON creation throws {\em InaccessibleObjectException}. If any WebSocket is a member field, it cannot be passed between the microservices 
as the socket IDs are environment specific.
\item Private member fields are exposed in JSON format.
\end{itemize}
JSON {\em  issues related to references:}
\begin{itemize}
\item References get lost~\cite{icws22}. The object passed from client to server API and back to the client after modification no more remains pointed by the original references at the client-side. Figure~\ref{jsonproblems}(a) shows an example from DayTrader where the object profile points to the object account via the field accountField on the client microservice. The object account is passed to the server microservice for the update of the field balance. The server microservice updates and returns the JSON of the updated object account. The JSON is deserialized on the client microservice, and a new object gets created. This deserialized JSON object should be pointed by the object account as well as all its aliases, namely profile.accountField in the client microservice. Finding all the aliases of the returned object is non-trivial.
\item Aliases are lost. In Figure~\ref{jsonproblems}(b), root.left and root.right are aliased in the client microservice. When it is serialized to JSON to be passed to the server, the JSON does not store the information that root.left and root.right are aliased. Therefore, when the server updates field root.right.name = "abc", field root.left.name is not updated to "abc" and remains "pqr". This happens because JSON does not save alias or address information. A correctly transferred and updated root object should have root.left.name = root.right.name = "abc". 
\item 'this' reference cannot be updated. Figure~\ref{jsonproblems}(c) is a code snippet from DayTrader where 'this' is passed as a parameter to an API TradeAction(.). If the state of 'this' is modified in the API, we need to update it on return of the API call. In other words, we need to execute {\em this = response.getThis();} on the client microservice after the API returns. However, 'this' cannot be updated.
\item The state of a modified object may not be reflected to the client. Even if the server returns the updated value of the object to the calling method in the client, the client cannot update the value of the object if the object is a parameter of the calling method. This is because if an object pointed by a Java parameter is updated, it is not reflected in its caller. The modifications are reflected in the caller only if fields of the Java parameter are updated. Figure~\ref{jsonproblems}(d) shows a monolith with method resetTrade(.), which calls method update(.), which further calls method setHoldingCount(.). Let us say microservices are made such that resetTrade(.) and update(.) are part of one microservice, and setHoldingCount(.) is part of another. In other words, the server microservice exposes setHoldingCount(.) as an API, which is called by update(.) in the client microservice. Method resetTrade(.) calls update(.), which passes runStats to API setHoldingCount(.). The API modifies runStats and needs to return the updated object to update(.). Method update(.) makes runStats point to this new updated object by executing {\em runStats = response.getRunStats();} in update(.) in the client microservice. However, runStats will continue to point to the old object in resetTrade(.), which is wrong; it should also be made to point to {\em response.getRunStats();}.  This is non-trivial to achieve.
\end{itemize}
JSON {\em  issues related to constructors and destructors:}
\begin{itemize}
\item Violates singleton property. A class can be made singleton by defining a private constructor. However, {fromJson(.)}, which deserializes JSON, violates access specifiers and can access private constructors. It creates new instances during deserialization, thereby violating singleton enforcement. In DayTrader, {\em MarketSummarySingleton} is annotated with @Singleton, which may be enforcing this. However, {fromJson(.)} would violate it.
\item A default constructor is written by the developer of the application to initialize member fields with default values. However, {fromJson(s)} calls the default constructor to create an object from JSON 's' and then overwrites the field values with the fields of 's'. This overwriting may lead to misguiding results. In Figure~\ref{jsonproblems}(e) default constructor reads all rows from a DB, writes to a file, and saves the number of rows read. Let us say, JSON of an object is {\em \{"recordsCount":10\}} and number of records in the DB are right now 100. When fromJson({\em \{"recordsCount":10\}}) is called at the server to deserialize the JSON, the default constructor is called, all 100 DB records are read into a file again, recordsCount is set to 100. After completing the operations of the default constructor, fromJson({\em \{"recordsCount":10\}}) wrongly overwrites value of recordsCount from 100 to 10. 
\end{itemize}

This section explained why objects could not be passed in JSON format. To overcome the challenges, we propose to create a unique ID for each object and communicate IDs instead of objects via APIs between the microservices. Our ID passing approach solves all the issues explained here.

%% file: Sections/communication.tex
\section{Communication using ID}
\label{communication}
In order to communicate data via IDs, we need to refactor the monolith and create wrapper classes. We explain the wrapper approach in Section~\ref{wrappersection} and then the ID passing approach in Section~\ref{sec:id}.

\subsection{Wrapper-Based Approach}
\label{wrappersection}
 Figure~\ref{wrapper}(b) shows the code changes required to convert a monolith to microservices architecture using JSON format for communication. Notice that the code becomes messy in the following sense: (i) the call to addItem(.) in the client microservice gets hidden in the creation of an API call, and (ii) the business logic on the server microservice gets hidden in the creation of the REST API. We propose a wrapper-based approach (Figure~\ref{fig:id}) to pack these steps into wrapper functions at the client and server-side, respectively.

 In the client microservice, class Populate is unmodified, and in the server microservice, members of class CatalogMgrServer are the same as the members of CatalogMgr. 
 New classes, namely, class CatalogMgr and class CatalogMgrWrapper, are added to the client and server microservices, respectively. These are used as wrappers on the client and server sides, respectively. Note that the class and function signature in class CatalogMgr of the client microservice has been kept unmodified, so that client code in class Populate remains unmodified. The steps required to expose a method as an API are put in the wrapper class in the server microservice. The steps required to make an API call are put in the wrapper class in the client microservice. We define addItem(.) in class CatalogMgr of the client microservice. Therefore, the method call addItem(.) in class Populate of the client microservice now calls the addItem(.) in the wrapper class CatalogMgr of the client microservice. This wrapper class makes a REST API call to the server microservice. We expose API addItem(.) in the wrapper class CatalogMgrWrapper of the server microservice. The API calls method addItem(.), which goes to the original method defined in the class CatalogMgrServer of the server microservice. Note that in the server microservice, we rename class CatalogMgr to CatalogMgrServer so that if the server microservice acts as a client and we need to add wrapper class CatalogMgr, it does not clash with the original class CatalogMgr.





\begin{figure*}[t]
\centering
\begin{tabular}{@{}c@{}}
\includegraphics[width=178mm]{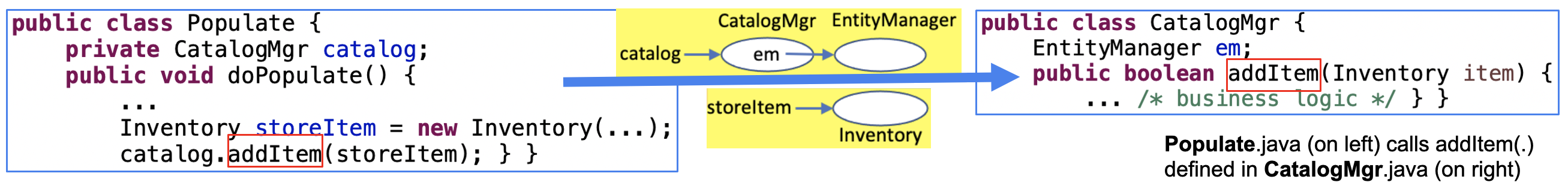} 
\\ \hline
\\[-2.5ex] \hline
\includegraphics[width=178mm]{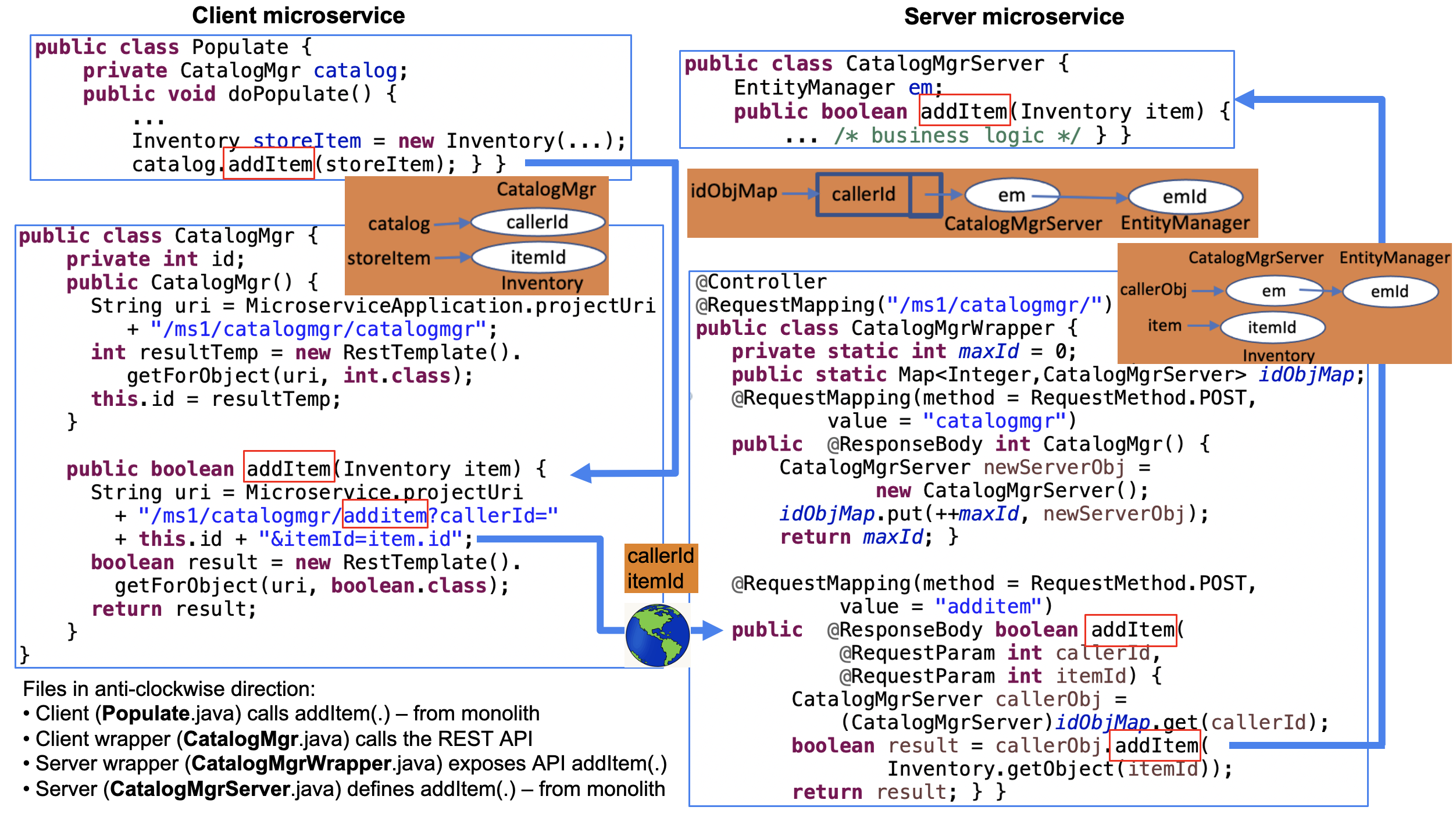} 
\end{tabular}
\caption{Plantsbywebsphere monolith (at the top). It is refactored into two microservices that communicate via APIs using ID approach.}
\label{fig:refactored}
\end{figure*}

Below are the advantages of creating wrapper functions.
\begin{itemize}
\item  \textit{Readability}: Business logic remains unmodified outside the wrapper functions, i.e., refactoring is non-intrusive. In Figure~\ref{fig:id}, call addItem(.) in class Populate of the client microservice and method definition of addItem(.) in class CatalogMgrServer of the server microservice remain unmodified.
\item \textit{Small code size}: 
Code duplication is reduced as calls to the same API correspond to the same wrapper function. For example, without the wrapper approach, if there are multiple calls to addItem(.) in class Populate, each call would have to be replaced with the REST API call. This would have caused code duplication. With the wrapper approach, each call to addItem(.) is unmodified and goes to the same addItem(.) of wrapper class CatalogMgr of the client microservice.   
\item \textit{Low local memory footprint}: No need for separate variables for each REST API call. Without the wrapper approach, if there are multiple calls to addItem(.) in class Populate, each call would be replaced with the REST API call; variables used for request, response, URI, etc. must be separate for each REST API call so that they do not lead to redefinition error. No extra local variables are required outside the wrapper in the wrapper approach.
\item \textit{Low maintenance cost}: Any modification to the API call in the refactored code is localized to the wrapper function. For example if there is any framework version update, the changes related to exposing the API would remain localized to the wrapper class CatalogMgrWrapper of the server microservice; it would not require changes in the original business logic in class CatalogMgr of the server microservice.
\item \textit{Refactoring is easier}: Without wrapper functions, method call has to be replaced with multiple statements for the REST call. This might cause a syntax error if the method call is part of another expression. Let us say, the method call is inside an if-construct: {\em if (addItem(...))}. We need to add the steps of making the REST API call before the if-construct and replace addItem(...) with {\em response.returnValue()}. Without the wrapper approach, automation would have become difficult.
\end{itemize}



Further, wrapper design handles overridden APIs automatically. When an overridden method is called by a superclass reference, which subclass or superclass method will be executed can be determined only at runtime. 
Let us extend the monolith example in Figure~\ref{wrapper}. Suppose class Z extends CatalogMgr, and class Z has its definition of addItem(.). The method call catalog.addItem(.) in class Populate would call CatalogMgr's addItem(.) or Z's addItem(.) depending on whether catalog holds the object of class CatalogMgr or Z. Whether catalog.addItem(.) should be replaced with {\em http://.../catalogmgr/addItem} or {\em http://.../z/addItem} cannot be determined at compile-time without the wrapper approach. However, with the wrapper approach, wrappers class CatalogMgr and class Z in the client microservice handle this, i.e., CatalogMgr's addItem(.) in the client microservice will have {\em http://.../catalogmgr/addItem}, and Z's addItem(.) in the client microservice will have {\em http://.../z/addItem}.


\subsection{ID-Passing Approach}
\label{sec:id}
Figure~\ref{fig:id} illustrates our ID-passing approach. Details of Figure~\ref{fig:id} are shown in Figure~\ref{fig:refactored}. The figures show snapshots of public application PlantsbyWebshpere, introduced in Section~\ref{introduction}.

On the server-side, the object is created and saved. Each object is {\em owned} by its creating microservice. A unique ID is created and returned to the client. A map from unique IDs to objects is saved in the server-side microservice. On the client-side, whenever an object needs to be created or accessed, the server is called to create or access the object, respectively. On creation, client gets a unique ID from the server and saves the ID. Client passes the unique ID in API calls to the server to access the object. To enable this, the field accesses are done via server APIs rather than direct statements.

In the client microservice, we add class CatalogMgr as a wrapper class. 
It contains an ID field, and no other fields of CatalogMgr are saved here. Wherever in the client microservice, CatalogMgr is instantiated, this wrapper class CatalogMgr constructor is called. It makes a REST API call to the server that creates an object of class CatalogMgr and returns the unique ID. Class CatalogMgr also contains a new definition of addItem(.), which makes REST API call to addItem(.) exposed by the server microservice.

 In the server microservice, we add class CatalogMgrWrapper, representing a wrapper class. It exposes the constructor of CatalogMgrServer as an API with signature {\em int CatalogMgr()}. It creates a new CatalogMgrServer, creates a unique ID, saves the ID and object in a map, and returns the ID to the client microservice. Class CatalogMgrWrapper also exposes addItem(.) as an API. This addItem(.) receives the ID of class CatalogMgrServer in parameter, retrieves the object from its map, and calls the original addItem(.) defined in CatalogMgrServer.  


If an application leads to too many API calls, a developer may keep local copies of the objects in all microservices, update them locally, and reflect the updates in the creator microservice periodically using their IDs.

\section{Communication using ID vs. JSON}
\label{comparison}

We elaborate below on how the three categories of issues in JSON passing listed in Section~\ref{existing} do not arise in the ID approach.
\begin{itemize}
\item JSON {\em issues related to access modifiers and data types}. Issues related to static fields, private members, and resource IDs do not arise in the ID approach as these are maintained on the microservice that created the object. Objects are not passed; only IDs are passed in this proposed approach. 
\item JSON {\em issues related to references}. Issues related to reflecting the state in the client microservice, passing aliases, and updating the state of 'this' do not arise in the ID approach. In the ID approach, any state update need not be reflected because objects are not passed; only their IDs are passed. The state of each object is maintained in the microservice that created it.
\item JSON {\em issues related to constructors and destructors}. This issue does not arise in the ID approach because objects are not re-created; their IDs are passed. Therefore, constructors/destructors are not called for the objects in API calls.
\end{itemize}

To further underpin our claim that ID-passing is decidedly superior over JSON-passing, we present the following advantages. 
\begin{itemize}
\item Less runtime: ID-passing does not require serialization and deserialization. 
\item Less refactoring of monolith: Unlike JSON, ID-passing does not require merging/unmerging of parameters, serialization/deserialization, and reflecting state updates.
\item Less memory and good data distribution: In the ID approach, objects are maintained only in the creator microservice. JSON approach saves all objects in the microservice that contains the monolith class with {\em main(.)}; thereby making that microservice a hotspot.
Let us say class {\em A} of microservice 1 contains the monolith class with {\em main(.)} and a reference, say {\em objB1}, of an object of class {\em B} of microservice 2. When an object of {\em B}, say {\em objB2}, is created in microservice 2, JSON approach saves its copy in {\em objB1} in microservice 1. Thus, microservice 1 saves the objects of both class {\em A} and {\em B}, and microservice 2 saves the object of class {\em B}. ID approach saves only the ID of {\em objB2} in microservice 1.
\end{itemize}


These points clearly bring out that ID-passing is decidedly superior to JSON-passing when migrating monoliths to microservices.

%% file: Sections/methodology.tex
\section{Reducing Data Communication}
\label{methodology}


   
  


Data communication via APIs can be reduced if the clustering of classes in the monolith application is done optimally. In other words, if functionally aligned methods are put in the same cluster, communication between the clusters can be reduced. However, dependencies across clusters are non-trivial to eliminate. These include field accesses, method calls, inheritance, interfaces, etc. We consider only field accesses and method calls in this paper.

\subsection{Intuition}
Steps 1 and 2 of Figure~\ref{system} illustrate dependencies between recommended clusters. Class {B} and  Class {C} from cluster2 access methods of class {A} from cluster1. Therefore a dependency edge is created from {B} to {A} and {C} to {A}. A finer level analysis shows that the top method of class {A} is accessed just by cluster2. This top method and its field are disconnected from other methods/fields of class A. Therefore, this top disconnected subgraph of class {A} can be relocated to cluster2 as class A' without much refactoring. However, all other disconnected subgraphs are being called by more than one cluster; therefore, we do not relocate them. For example, bottom subgraph of class A is being called by clusters 2 and 3. Similarly, bottom subgraph of class C is called by clusters 1 and 3.


Reconstructing a class that implements a fraction of the functionality of another class such that it introduces no new errors is an undecidable problem~\cite{csur17} due to presence of pointers~\cite{ismm17}. Our approach focuses on identifying disconnected components (direct field and method calls) that can be moved with minimal side effects. 

 \begin{figure*}[t!]
 \centering
 \includegraphics[width=7in, trim = {0cm 3cm 0cm 0cm}]{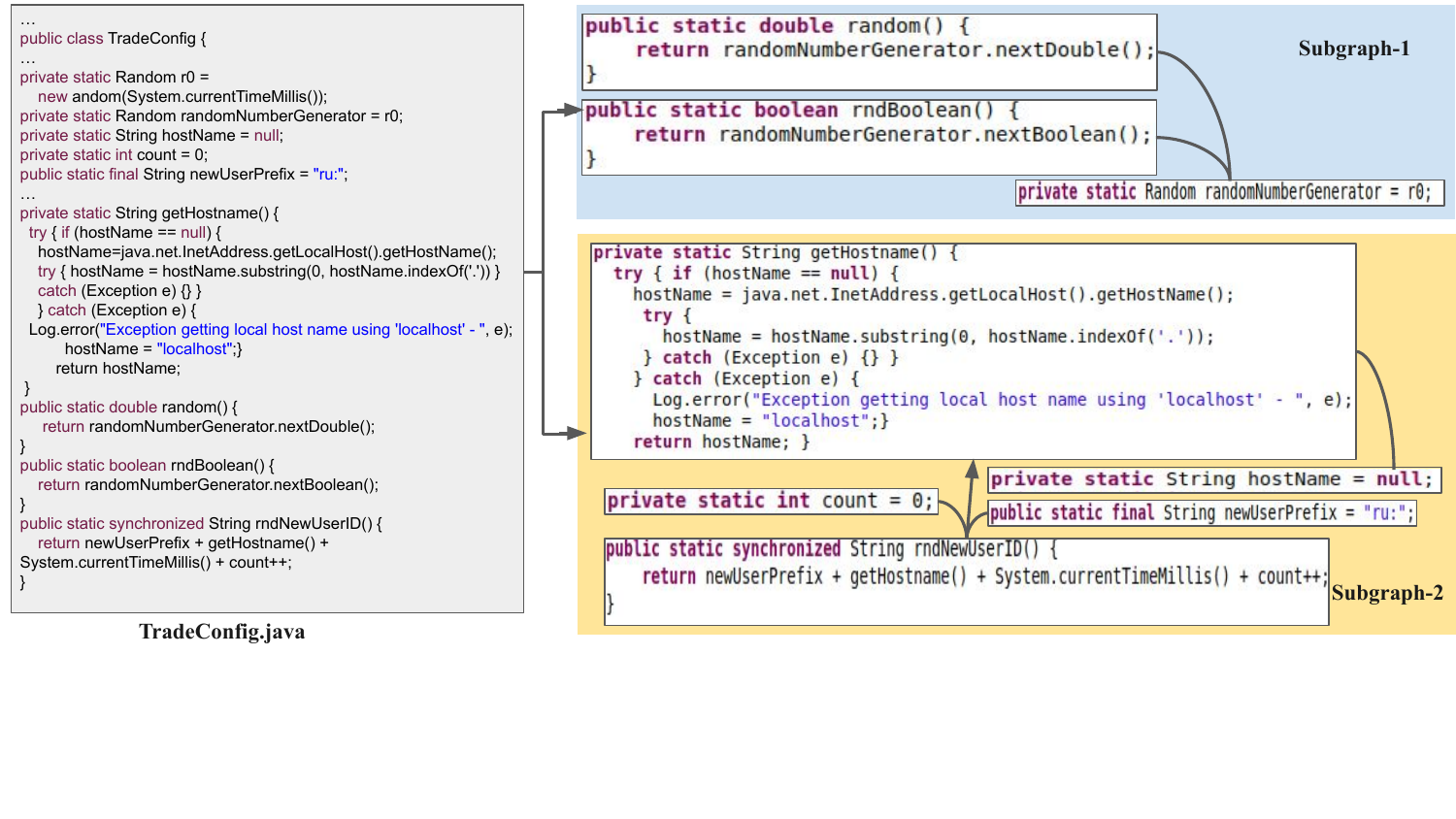}
 \caption{Representation of two disconnected subgraphs created for TradeConfig.java in the DayTrader application. }
 \label{subgraph}
 \end{figure*}

\subsection{Relocating Disconnected Subgraphs} \label{disconnected-subgraph} 
A subgraph with nodes as methods and fields in a class, is disconnected from other subgraphs if there is no path connecting any nodes of two subgraphs. Disconnected subgraphs are basically functionally dissociated code blocks in the same class. 

To build disconnected subgraphs from method calls and field accesses, we construct a call graph such that (i) nodes represent methods and fields and  (ii) edge from node {\em n1} to {\em n2} exists if method {\em n1} accesses method {\em n2} or field {\em n2}. Then we find disconnected subgraphs from the call graph.
Once a disconnected subgraph is relocated to the target cluster, the caller code is modified accordingly. 

Figure~\ref{subgraph} illustrates two disconnected subgraphs in class TradeConfig.java of DayTrader~\cite{daytrader}. Subgraph2 contains method rndNewUserID(.), which calls getHostname(.) and they use fields count, hostName, and newUserPrefix. Subgraph2 is called only by class TradeScenarioServlet in another cluster. Relocating this subgraph from TradeConfig to the cluster containing TradeScenarioServlet will make the clusters more functionally aligned. Moreover, relocating this subgraph will make the call to subgraph2 as a local call in cluster TradeScenarioServlet rather than an API call.

Subgraph1 is not a candidate for relocation as explained below. It contains methods random(.) and rndBoolean(.) connected via field randomNumberGenerator. Since rndBoolean(.) is being called by only OrderDataBean, it could have been relocated to the cluster containing OrderDataBean. However, random(.) is being called by several classes. Therefore, it cannot be moved. Due to the field dependency between random(.) and rndBoolean(.), we do not relocate either of the methods.

%% file: Sections/exposingApis.tex
\section{Exposing Methods as APIs}
\label{exposingapis}
There are multiple ways of communication between microservices, including remote procedure calls, REST APIs exposed using Springboot~\cite{spring-boot}, JAX-RS~\cite{jaxrs}, and other such frameworks for Java applications. Although we discuss REST APIs, our approach can be applied to other approaches too. Further, we demonstrate the API generation process using Springboot~\cite{spring-boot}, but it can be replaced with any other framework.
Below we explain the steps to expose a method as an API and the steps to convert a call into REST API call. 
These changes make the code bit hard to maintain. Therefore, in Section~\ref{wrappersection}, we have proposed to create wrappers on client and server side, which would encapsulate the communication logic code. Since we do not modify the original business logic, coding standards like statelessness, and concurrency, if present, are carried forward to the generated microservices. 

 

For each candidate microservice, we automate the following steps to create a deployable project. We automatically generate swagger file, create repositories for each microservice, create a @SpringBootApplication starter class, copy programs and interfaces to appropriate repositories, import required packages, and create build configuration files like pom.xml or gradle.


We automate the following steps required on the server-side to expose APIs using Springboot. We automatically add Springboot annotations like @Controller, @RequestMapping, @ResponseBody, @RequestParam, convert arguments/parameters to a format that can be shared between microservices (Section~\ref{communication}), update API signature, unmerge non-primitive parameters, return HTTP and Exception status code, and encapsulate method body in try-catch.


We automate the following steps required on the client-side to call APIs.
We automatically create URI with query parameters, create \textit{RestTemplate()} and HTTP calls such as get/post, etc., merge multiple non-primitive arguments, and format the arguments so that they can be passed to other microservices (Section~\ref{communication}), and process the response of the API.



%% file: Sections/discussion.tex
\section{Coding Design Considerations}
\label{discuss}


While transforming a monolith into microservices, we came across some coding design choices that were good for the monolith but non-conventional for microservices. These anti-patterns create an issue with JSON-passing between microservices (see Section~\ref{communication}). 
For a good microservices design, we list below some coding designs. 
If these coding designs are followed, issues with JSON-passing will not appear.
The list of anti-patterns may not be complete and are only suggestions to the user for remediations.
\begin{itemize}
\item Non-primitive objects containing static member fields should not be shared between microservices because these are difficult to pass (Section~\ref{existing}, Figure~\ref{jsonproblems}(a)). DayTrader classes {\em TradeConfig} and {\em TradeAction} have several static member fields.
\item State of the arguments should not be modified. Section~\ref{existing} using Figure~\ref{jsonproblems}(b) shows that reflecting the state to the client microservice is difficult. 
\item References should not be maintained in API-based workflow (Section~\ref{existing}, Figures~\ref{jsonproblems}(b) and (c)). 
\item Simple getters/setters should not be marked as APIs as they increase interactions between microservices. There are several such getters/setters in DayTrader.
\item External resources like files, sockets, tables, queues should be referenced via some external configuration accessible to the microservices. E.g., file path or dataset name, IP and port for the socket, DB connection parameters and fully qualified table name, queue manager parameters and queue name. There are several external resources like {\em HttpServletRequest}, {\em ServletConfig}, {\em ServletOutputStream}, {\em ReadListenerImpl}, and others that are passed via methods in DayTrader.
\end{itemize}


%% file: Sections/experiments.tex
\section{Implementation and Measurements}
\label{experiments}

\begin{figure}
\centering
\resizebox{\linewidth}{!}{%
\scriptsize
\begin{tabular}{|l|l|l|l|l|l|}
\hline
\textbf{Dataset} & \textbf{Clusters} & \textbf{Classes} & \textbf{Methods} & \textbf{Fields}  & \textbf{Time (mins.)}  \\
\hline
DayTrader & 6 & 111 & 952 & 530 & 8.19 + 13.12 \\
\hline
PBW  & 5 & 36 & 424 & 276 & 0.90 + 6.64 \\
\hline
Acme-Air & 4  & 38 & 196 & 117 & 1.18 + 4.46 \\
\hline
Petclinic  & 4 & 37 & 138 & 57 & 1.18 + 2.90 \\
\hline
Mayocat  & 7 & 667 & 3042 & 1276 & 19.96 + 63.00 \\
\hline

\end{tabular}
}

\caption{Monolith applications. The execution time is for function isolation and refactoring using ID approach.}
\label{tab:dataset}

    \centering
    \begin{tabular}{|c|}
    \hline
    \includegraphics[width=83mm]{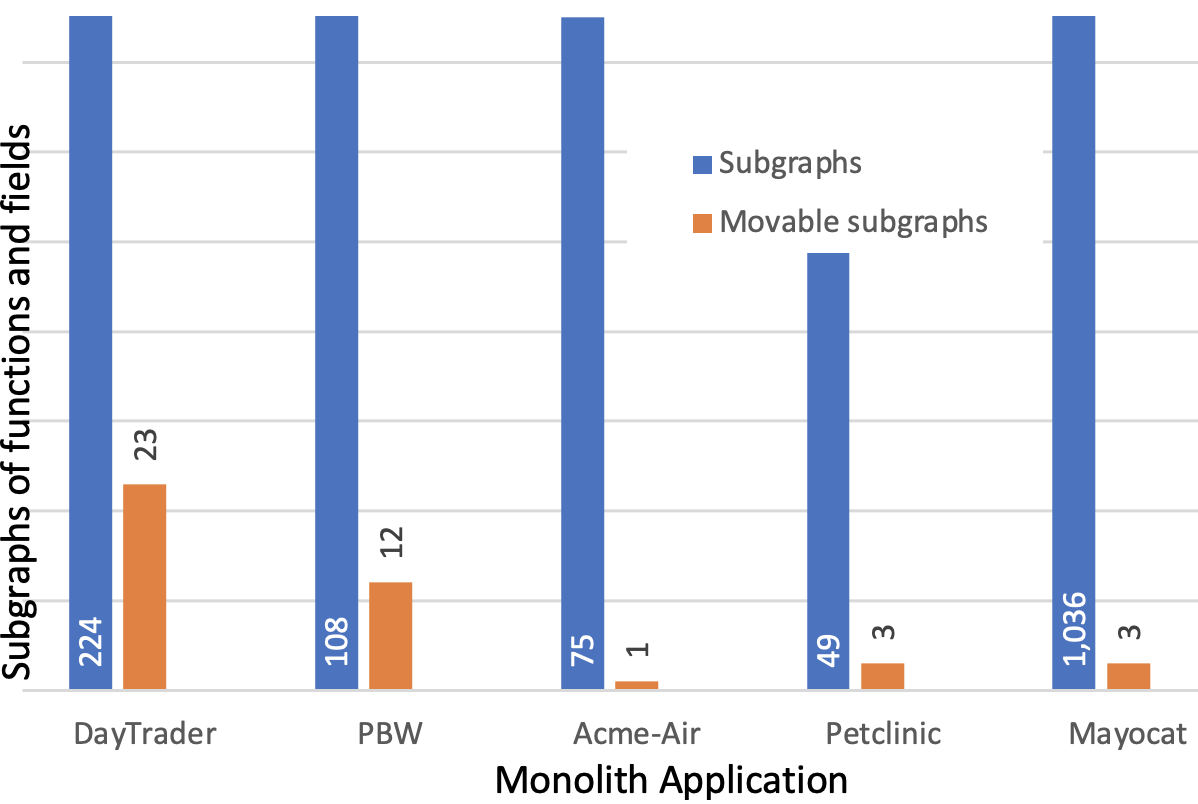} 
    \footnotesize \hspace{-8.6cm} \bf (a) \hspace{8.1cm} \\ \hline
    \includegraphics[width=83mm]{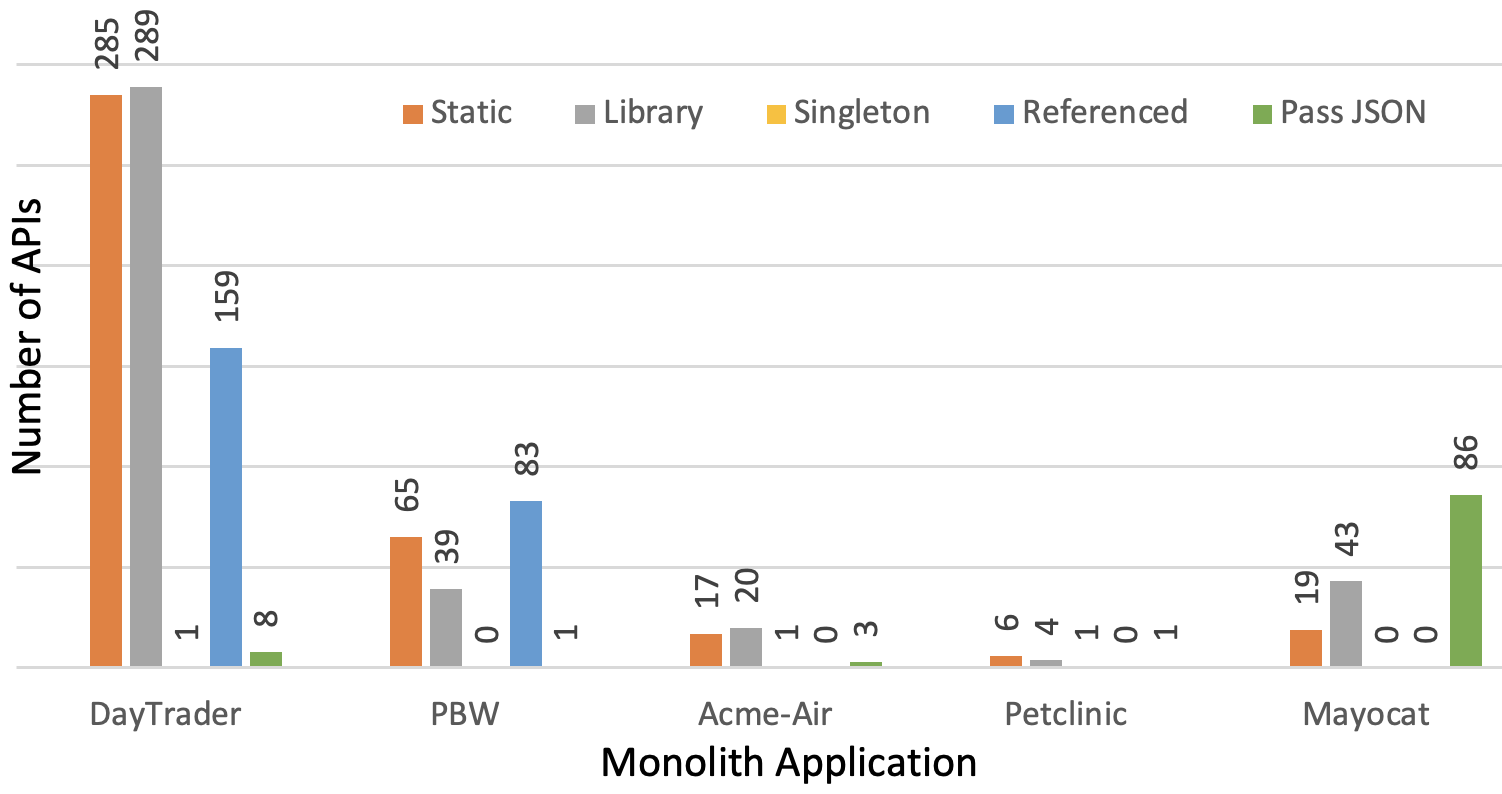} 
    \footnotesize \hspace{-8.6cm} \bf (b) \hspace{8.1cm}
    \\ \hline
    \includegraphics[width=83mm]{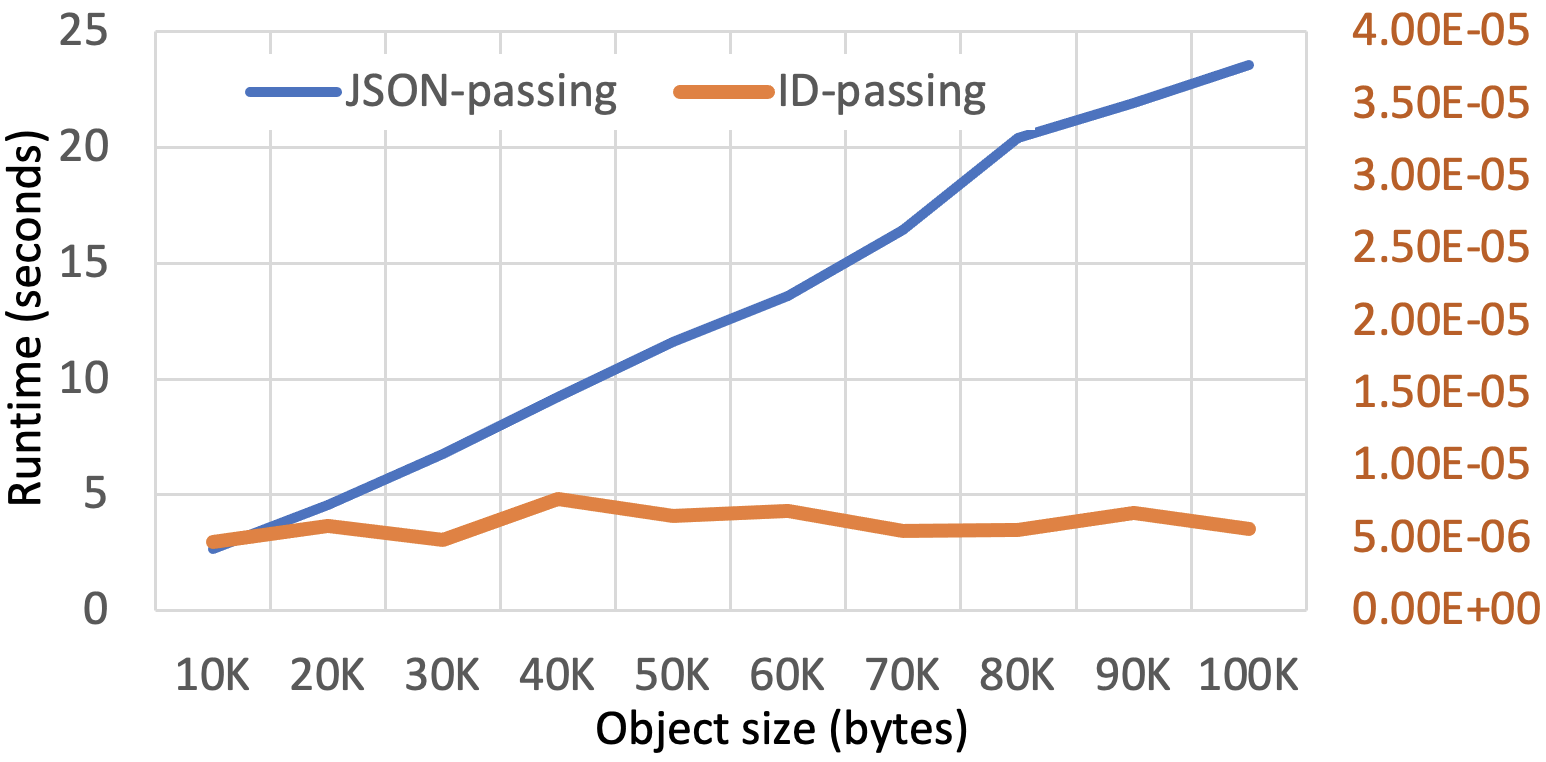} 
    \footnotesize \hspace{-8.6cm} \bf (c) \hspace{8.1cm} \\ \hline
    \includegraphics[width=83mm]{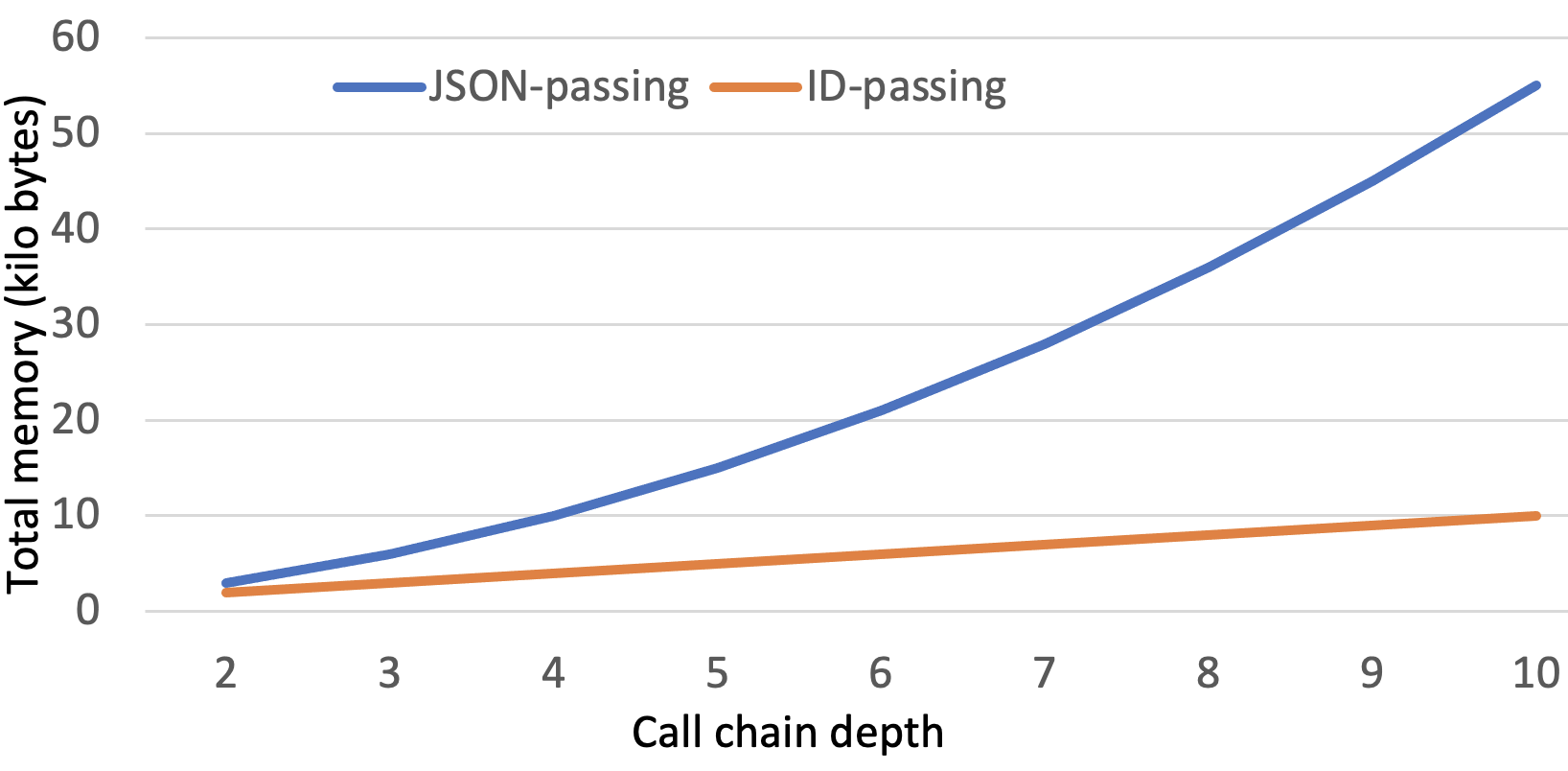}
    \footnotesize \hspace{-8.6cm} \bf (d) \hspace{8.1cm} \\ \hline
    \end{tabular}
    \caption{(a) API reduction  (b)  Number of APIs which require ID vs. JSON (c) Runtime of JSON vs. ID (d) Memory usage of JSON vs. ID }
    \label{fig:experiments}
\end{figure}

In this section, we present experiments on reduction in data communication (Section~\ref{sec:exp1}) and coverage and performance comparison (Section~\ref{sec:exp2}). We also present a qualitative study using a survey (Section~\ref{sec:exp3}). Finally, we compare our automation with manually refactored PetClinic monolith (Section~\ref{sec:exp4}).

\subsection{Environment and Benchmarks}
Our experiments for function isolation (steps 1 and 2 of Figure~\ref{system}) and code refactoring (step 3 of Figure~\ref{system}) were conducted respectively on laptops of the following configurations: Intel x86/64 i7-4510U up to 3.10GHz, Ubuntu 18.04OS, 8GB RAM and macOS Big Sur 11.4, 2.8 GHz Quad-Core Intel i7, 16 GB RAM.

 We studied our approach on five publicly-available web-based monolithic java applications, viz. Daytrader~\cite{daytrader},
 Acme-Air~\cite{acmeair}, Plantsbywebsphere~\cite{pbw}, Petclinic~\cite{spring-petclinic} and Mayocat~\cite{mayocat} (Figure~\ref{tab:dataset}). 
 These applications are commonly studied for monolith to microservices migration task~\cite{kalia2020mono2micro, desai2021graph}.  Our implementation and results of the experiments for the five applications are available on Github~\cite{res}.
 
\subsection{Measuring Reduction in Data Communication}
\label{sec:exp1}
For building graphs with direct calls (without pointers) we use tools like JavaParser~\cite{javaparser} and Soot~\cite{soot}. Our framework takes as input candidate clusters, i.e., groups of classes (step 0 of Figure~\ref{system}). For this work, we leverage existing cluster recommendations~\cite{desai2021graph}, but our work can consume any cluster recommendation tool output. 
Then using our approach of function isolation (Section~\ref{methodology}), we measure the number of relocatable disconnected subgraphs. 
These are the subgraphs whose methods need not be exposed as APIs. Figure~\ref{fig:experiments}(a) shows that DayTrader has 224 disconnected subgraphs; 23 can be relocated, i.e., their methods need not be exposed as APIs. DayTrader monolith consisting of 111 classes, had six candidate clusters of these classes. Using our function isolation algorithm, we find that its  952 methods and 530 fields have 224 disconnected subgraphs. Our implementation detected that out of these, 23 disconnected subgraphs can be relocated~\cite{res}. This naturally reduces the dependencies between the clusters. For example, a disconnected subgraph in class {\em KeySequenceDirect} from recommended cluster {\em MDBQueue\_service} with two methods: \textit{getNextID(.)} and \textit{allocNewBlock(.)},  and three field accesses: \textit{getKeyForUpdateSQL}, \textit{updateKeyValueSQL} and \textit{createKeySQL} can be relocated to microservice {\em SessionEntity\_service} since the disconnected subgraph is only called by class {\em TradeDirect} that belongs to this cluster. Relocating this subgraph to its more functionally aligned cluster, reduces the number of APIs that need to be exposed.

\subsection{Coverage and Performance Comparison}
\label{sec:exp2}
We empirically evaluate the coverage, i.e., the cases where JSON loses information. These cases are also mentioned as legends in Figure~\ref{fig:experiments}(b).
(i) {\em Static}: Caller, parameter, or return object contains a static member field.
(ii) {\em Library}: Caller contains a member field of library type, or parameter or return object is library type.
(iii) {\em Singleton}: Caller, parameter, or return object’s constructor is private or singleton.
(iv) {\em Referenced}: Caller, parameter, or return object or their field is referenced by another object; we detect this using type analysis~\cite{csur17}. 
Legend {\em Pass JSON} denotes APIs where caller, parameter, and return objects do contain the above constructs. Thus, in these APIs, JSON can be used for data transfer. This count is very small in all the applications. This clearly establishes that ID is required over JSON in most of the APIs.

Figure~\ref{fig:experiments}(c) shows that runtime using JSON vs. ID for different object sizes is around $\mathcal{O}(n)$ vs. $\mathcal{O}(1)$. This is because runtime for JSON includes serialization and deserialization. 
 
Figure~\ref{fig:experiments}(d) shows the total memory usage in all microservices for different call chain depths. JSON approach creates as many copies as the call chain depth. ID approach keeps only one copy of each object (Section~\ref{comparison}). If there are {\em n} microservices with one class each, the longest call chain depth possible is {\em n}. Let the average object size be 1KB. In JSON, there will be one copy of $n^{th}$ object, two copies of $(n-1)^{th}$ object, and so on. These add up to {\em n(n+1)/2} KB of memory. In ID, there will be a single copy of each object which adds up to $n$ KB of memory. Thus, memory requirements of JSON vs. ID for different call chain depths is $\mathcal{O}(n^2)$ vs. $\mathcal{O}(n)$. These establish the superiority of ID over JSON approach for data transfer. 

\begin{figure}
\centering

\begin{tabular}{@{}l@{}}
\begin{tabular}{|@{}l@{}|}
\hline
\includegraphics[width=65mm]{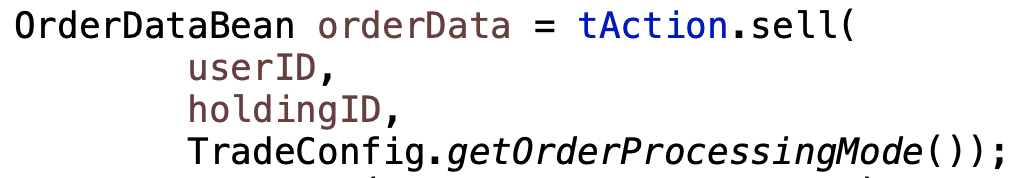} \\ \hline
\footnotesize TradeServletAction.java calls sell()
 \\ \hline \hline
\includegraphics[width=90mm]{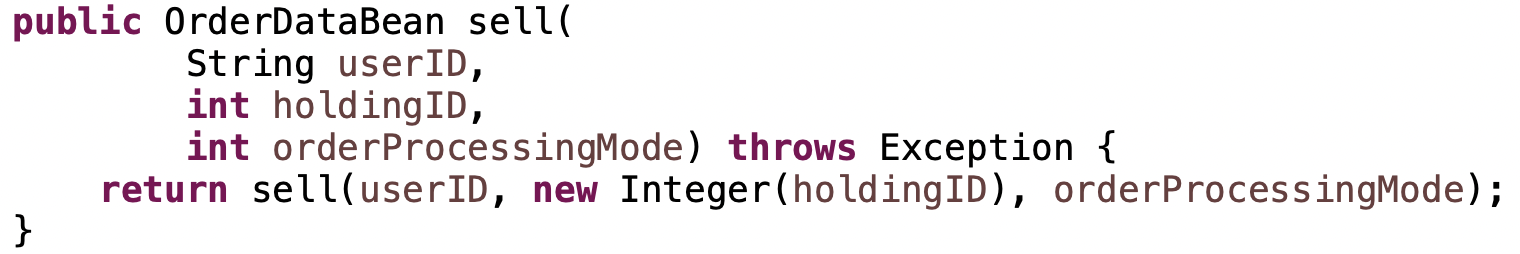}\\ \hline
\footnotesize TradeAction.java defines sell() \\ \hline
\end{tabular} \\
{\bf \small (a) Monolith code} (Input). \\ \\

\begin{tabular}{|@{}l@{}|}
\hline
\includegraphics[width=70mm]{images/TradeServletActionServer.png} \\ \hline
\footnotesize Client (TradeServletActionServer.java) calls sell() -- unmodified from monolith
 \\ \hline \hline
\includegraphics[width=85mm]{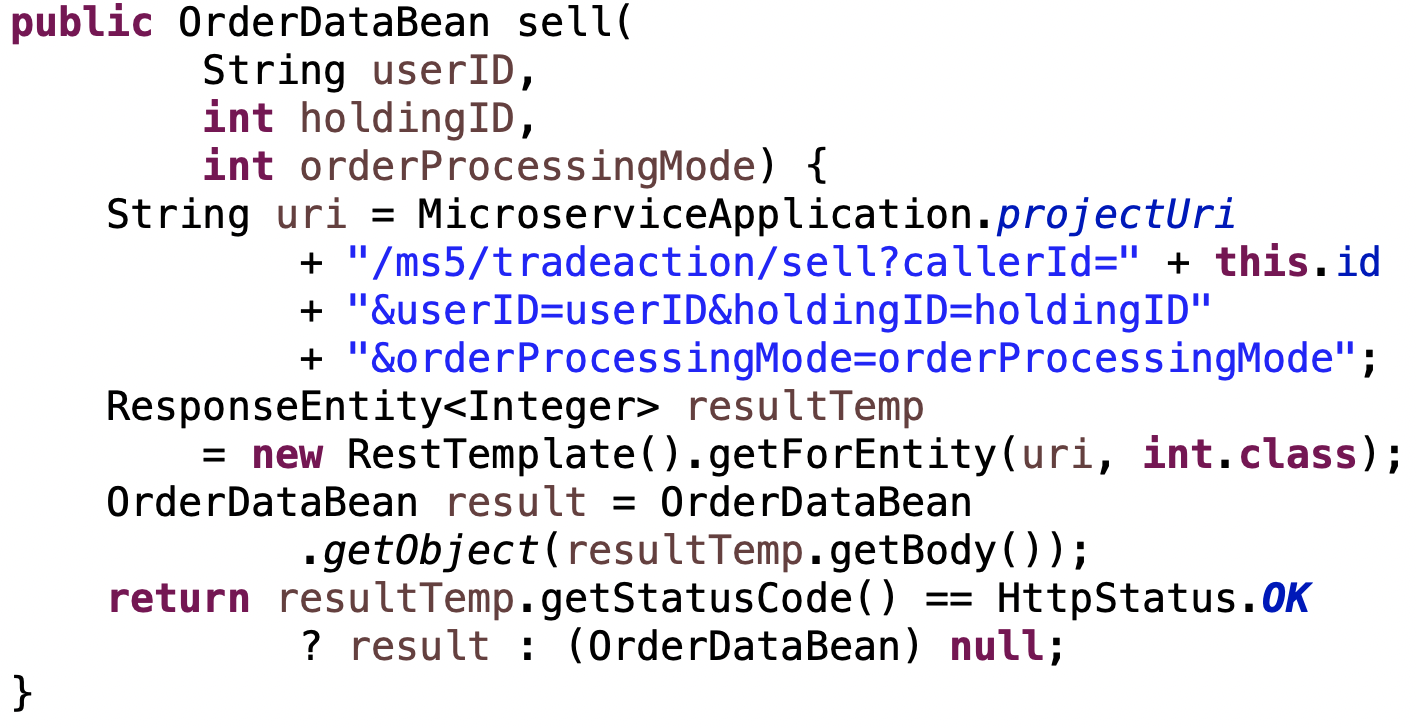} \\ \hline
\footnotesize Client wrapper (TradeAction.java) calls the REST API \\
\hline \hline
\includegraphics[width=85mm]{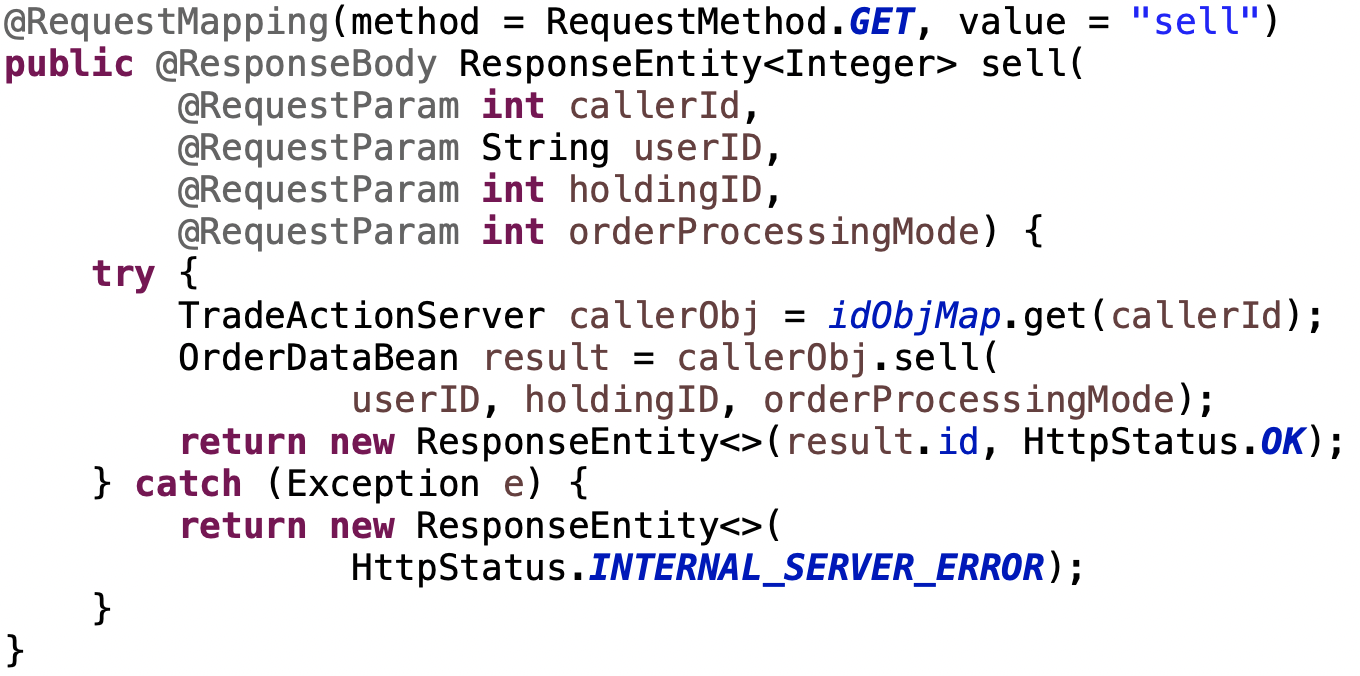} \\ \hline
\footnotesize Server wrapper (TradeActionWrapper.java) exposes REST API sell() \\ \hline \hline
\includegraphics[width=90mm]{images/TradeActionServer.png}\\ \hline
\footnotesize Server (TradeActionServer.java) defines sell() -- unmodified from monolith \\ \hline
\end{tabular} \\
{\bf \small (b) Refactored client-server microservices} (Output).
\end{tabular}
\caption{Original DayTrader monolith (input) and its refactoring (output from our tool).}
\label{refactored}
\end{figure}

A sample from monoliths PlantsByWebshere and DayTrader have been refactored to microservices and shown in Figures~\ref{fig:refactored} and \ref{refactored}, respectively.

\subsection{Threats to Validity}
\label{threats}
In theory, our function isolation and refactoring preserve the behaviour of the application. However, the following are some threats to achieving them.

\emph{External threat: }
Since our approach relies on the call-graph analysis to identify the relocatable candidates, the complexity may increase with the size of the monolith. 

\emph{Internal threat: }
There may be various connections and dependencies between classes defining a monolith. We limit our focus to field accesses and direct method calls. The accuracy of our tool depends on cluster recommendation~\cite{desai2021graph} considered as input.

%

\subsection{Soundness and Validation of Results}
Since there is no standard baseline work to compare, we validated our results by manually checking that the identified disconnected subgraphs were indeed disconnected. Further, we validated the final microservices by checking for compilation errors in generated code and executing its APIs. Our testing and quality analysis team also performed black box testing on the refactored application. 


%% file: Sections/study.tex
\subsection{Qualitative Study}
\label{sec:exp3}
To study the quality of the refactored microservices and their benefits to the developers, we conducted a survey through two software engineers and one researcher with program analysis background. On an average, the participants had an industrial experience of 13 years in different software engineering roles. All three participants had prior working experience with Java programming language and microservices. It took participants around four weeks to understand Daytrader along with their daily work. The participants were provided with a high-level summary of our work including the original application and our refactoring; no further instructions were given to avoid bias. The effort needed to understand an application along with necessary skill sets to assess output limited our study to few participants and datasets.

\subsubsection{Study Instructions}
With each participant we explained our code refactoring approach using examples from DayTrader and Plantsbywebsphere and all the figures in this paper. They examined the codes, cross questioned us, manually refactored the code on their own, and then answered the following questions :

\begin{itemize}
\item How much development effort does this refactored output reduce? a) Less ($<$20\%) b) Moderate (20–40\%) c) 40-60\% (Significant) d) $>$60\% (Completely)
\item How much would you modify before deploying the refactored work? a) No change (Use the code as such) b) Low (change $<$20\%) c) Moderate (20-40\%) d) Significantly ($>$40\%)
\item Please list the type of changes you would do.
\item Optionally, please explain your inputs.
\end{itemize}

\subsubsection{Results}
All three participants said that compared to manual refactoring, automation will save their delivery time in generating microservice projects and APIs by 40-60\% and reduce human errors~\cite{soeny}. A participant mentioned that he avoids making manual changes and relies on automatic refactoring to prevent the introduction of new errors. The other two participants said they would make moderate changes as they felt the number of APIs are too many; they may group the exposed APIs. All three participants mentioned that they would introduce the following changes 1) Modify API and variable names, 2) improve candidate clusters by introducing new endpoints more aligned with their business domain model, 3) remediate static fields, etc. Note that (2) is out of our scope as we do not perform clustering but expose functions of the given clusters as REST APIs. This study shows that our output accelerates their development task but more business-centric changes are still required.

\subsection{Manual Refactoring of PetClinic Monolith}
\label{sec:exp4}
There is no baseline to compare our work with. However, manual refactoring of a monolith to microservices has been attempted, for example, for Petclinic~\cite{ms-petclinic}. Like our approach, it also passes the IDs of objects for communication via APIs. For example, method {\em processCreationForm(.)} which passes non-primitive objects in a monolith, passes the ID of an object in the microservice design. This refactoring was easy to achieve in Petclinic because the monolith design already saved each object's IDs in repositories that could be directly used in the microservice design.

Nonetheless, we automatically introduce IDs since they may not be always present in the monolith. Moreover, we save objects in the memory local to each microservice, which is better than saving in central Database (as done by microservices Petclinic~\cite{ms-petclinic}). It is better because it is not exposed to all microservices and is faster than Database access.

\subsection{Summary}
Overall, we demonstrated that ID-passing is decidedly superior to JSON-passing as the former preserves application behaviour and has lower order of runtime and memory complexities. Further, to reduce communication across microservices, we relocated disconnected subgraphs of methods and fields to their more functionally aligned microservices; we are able to reduce up to 10\% of the APIs. Finally, our method reduces 40-60\% of the manual refactoring efforts, making it an important building block in automation projects.

%% file: Sections/related.tex
\section{Related work}
\label{related}
Literature has studied how to cluster classes into microservices. However, there is very little work on how to refactor the monolith code into deployable microservices automatically. Literature neither highlights nor handles the plethora of issues in JSON passing for communication via APIs. In our work, we address this critical research gap by taking candidate microservices, i.e., clusters, as inputs, reducing communication between microservices wherever possible, and then exposing the remaining methods as REST APIs to enable communication between microservices. We are not aware of any baseline work on automated refactoring to compare against to the best of our knowledge.

\subsection{Migrating from Monolith to Microservices}
Migration from monolith to deployable microservices is a big challenge. There are a lot of manual approaches proposed for developers to follow~\cite{kecskemeti2016entice, 9211252}.
Ren et al.~\cite{ren2018migrating} propose the creation of detailed Data Flow Diagrams (DFDs) for the different business use cases with application owners and then decomposition of DFDs to recommend microservices. Abdullah et al.~\cite{abdullah2019unsupervised} propose an approach to group URIs based on the response time collected from access logs. Also, monolith application is copied onto different virtual machines, serving specific endpoints to mimic microservices. One of the main reasons automated migration remains under-explored is that decoupling can introduce new errors. Therefore, many studies have focused on improving the accuracy of functional boundaries in the monoliths to minimize developer effort in taking the recommendations into deployable microservices.

Several work~\cite{desai2021graph, mathai2021monolith, tzerpos2000accd, harman2002new, mazlami-icws-2017, mahouachi2018search,  mancoridis1999bunch} leverage syntactic relationships to identify clusters as candidate microservices that perform well on standard clustering metrics like cohesion, coupling, number of modules, amount of changes, etc. In addition to finding functional clusters, Agarwal et al. ~\cite{agarwal2021monolith} introduced non-functional clusters like utilities, dead code, and refactorable programs that need developer attention. But none of the work discusses steps to take recommendations for deployable microservices. 

Kanvar et al.~\cite{icws22} discuss about information loss while transferring data objects containing pointers as JSON between microservices. They also propose pointer swizzling to solve this problem. However, they do not discuss or solve the plethora of other issues in using JSON format for communication. They still use JSON for communication via APIs which changes application behaviour. In this work, we discuss the new challenges and propose ID approach based solution.

\subsection{Distribution of Monolithic Application}
Multiple approaches exist for distribution of applications~\cite{10.1145/1555392.1555394, philippsen1997javaparty, zhang2012towards, mcgachey2011class, fuad2002adjava}; however, unlike our refactoring of monolith to microservices, they are not completely automated. J-Orchestra~\cite{10.1145/1555392.1555394, philippsen1997javaparty} is one of the early work on distributing stand-alone Java applications. They provide interface for the user to specify programs and system resources and generate RMI
stubs/skeletons to create a client-server model. DPartner~\cite{zhang2012towards}
identifies programs that should be retained or moved based on its usage of native libraries (e.g., Windows DLL) and then automatically create RMI stubs/skeletons for communication. RuggedJ~\cite{mcgachey2011class}, JavaParty~\cite{philippsen1997javaparty} and AdJava~\cite{fuad2002adjava} transform programs into custom object model that can scale on-demand to effectively harness the compute power. But none of these reduce the communication between microservices. 

%% file: Sections/conclusion.tex
\section{Conclusions and future work}
\label{conclusion}

In this work, we highlighted the existing gap in the literature to take the candidate microservices, i.e., clusters of programs to deployable microservices. We discussed the need for automatic code refactoring, which includes (a) exposing endpoints, (b) converting method calls to REST API calls, and (c) communicating data and resources between microservices. 
We identified and categorized the plethora of issues in using JSON for communication via APIs between microservices. There is very little work on this in literature. We propose an ID-passing approach for communication via APIs between microservices. Further, we propose an approach to reduce communication between microservices. We do this by constructing disconnected subgraphs of methods and fields and moving them to their more functionally aligned microservices. 
We demonstrated that ID-passing is decidedly superior to JSON-passing as the former preserves application behaviour and has lower order of runtime and memory complexities. 
Our qualitative study shows that this automation reduces 40-60\% of the manual refactoring efforts.

This work opens up interesting directions (a) remediate code that does not adhere to standard microservice coding patterns and (b) refactor interactions of programs with micro DBs.
In future, we plan to incorporate a wider set of inter-class and intra-class dependencies, such as interfaces and inheritance. We also plan to leverage existing business domain models like BIAN~\cite{bian}, TM Forum~\cite{tmforum} to determine the possible APIs to expose and their signatures.